\documentclass[preprint,showpacs,aps,fleqn]{revtex4}%
\usepackage{graphicx}
\usepackage{psfrag}
\usepackage{amsmath}
\usepackage{subfigure}

\begin{document}
\def\be{\begin{eqnarray}}
\def\ee{\end{eqnarray}}
\newcommand{\nn}{\nonumber}
\def\mpcomm#1{\nextline\strut\kern-6em{\tt MP COMMENT => \ #1}\nextline}
\def\nextline{\hfill\break}
\newcommand{\rf}[1]{(\ref{#1})}
\newcommand{\beq}{\begin{equation}}
\newcommand{\eeq}{\end{equation}}
\newcommand{\bea}{\begin{eqnarray}}
\newcommand{\eea}{\end{eqnarray}}
\newcommand{\pint}{-\hspace{-11pt}\int_{-\infty}^\infty }
\newcommand{\Pint}{-\hspace{-13pt}\int_{-\infty}^\infty }
\newcommand{\nint}{\int_{\infty}^{\infty}}
\renewcommand{\vec}[1]{\boldsymbol{#1}}

\title{Classical Strongly Coupled QGP: \\
VI. Structure Factors}
\author{Sungtae Cho}
\email{scho@grad.physics.sunysb.edu}
\author{Ismail Zahed}
\email{zahed@zahed.physics.sunysb.edu}
\affiliation{Department of Physics and Astronomy\\
State University of New York, Stony Brook, NY 11794-3800}

{\begin{abstract}
We show that the classical and strongly coupled QGP (cQGP)
is characterized by a multiple of structure factors that
obey generalized Orstein-Zernicke equations. We use the canonical
partition function and its associated density functional to derive
analytical equations for the density  and charge monopole structure factors
for arbitrary values of $\Gamma=V/K$, the ratio of the mean potential
to Coulomb energy. The results are compared with SU(2) molecular dynamics
simulations.
\end{abstract}
}

 \maketitle

\section{Introduction}

High temperature QCD is expected to asymptote a weakly coupled
Coulomb plasma albeit with still strong infrared divergences. The
latters cause its magnetic sector to be non-perturbative at all
temperatures. At intermediate temperatures of relevance to
heavy-ion collider experiments, the electric sector is believed to
be strongly coupled.

Recently, Shuryak and Zahed~\cite{SZ_newqgp} have suggested that
certain aspects of the quark-gluon plasma in range of temperatures
$(1-3)\,T_c$ can be understood by a stronger Coulomb interaction
causing persistent correlations in singlet and colored channels.
As a result the quark and gluon plasma is more a liquid than a gas
at intermediate temperatures. A liquid plasma should exhibit
shorter mean-free paths and stronger color dissipation, both of
which are supported by the current experiments at
RHIC~\cite{hydro}.

To help understand transport and dissipation in the strongly
coupled quark gluon plasma, a classical model of the colored
plasma was suggested in~\cite{gelmanetal}. The model consists of
massive quarks and gluons interacting via classical colored
Coulomb interactions. The color is assumed classical with all
equations of motion following from Poisson brackets. For the SU(2)
version both molecular dynamics simulations~\cite{gelmanetal} and
bulk thermodynamics were recently presented~\cite{cho&zahed2}
including simulations of the energy loss of heavy
quarks~\cite{dusling&zahed}.

In this paper we follow up on our recent equilibrium analysis of
the bulk thermodynamics~\cite{cho&zahed2} to the static structure
factors. In section 2 we define the energy functional for the cQGP.
In section 3 we derive generalized Ornstein-Zernicke equations
for the pair correlation functions. In section 4 we show that the cQGP
supports multiple structure factors that measure a variety of colored
correlations. Each structure factor obeys a generalized Ornstein-Zernicke
equation. In section 5 we introduce the Debye-Huckel-hole potential
for the cQGP. In section 6,  we use Debye charging process to derive
analytical expressions for the lowest two structure factors in the cQGP
for arbitrary $\Gamma$. In section 7, we construct numerically the
lowest two structure factors using molecular dynamics simulations
and compare them with our analytical results for values of $\Gamma$
in the liquid phase. Our conclusions are in section 8. Appendix A
is added to streamline our conventions for the SU(2) color charges.

\section{Free Energy Functional}

\renewcommand{\theequation}{II.\arabic{equation}}
\setcounter{equation}{0}

We consider the canonical partition function of a single species,
either quarks and gluons, at finite temperature $1/\beta=T$ and in
the presence of an external scalar source $\psi$

\begin{eqnarray}
\mathcal{Z}_N[\psi] & &=\frac{1}{N!}\int\prod_i\frac{d\vec r_i
dQ_i}{\lambda^{3}}\exp{\Big(\beta\int d\vec r dQ n(\vec r,\vec Q
)\psi(\vec r,\vec Q)\Big)}
\nonumber \\
& & \times \exp{\Big(-\frac{\beta}{2}\frac{g^2}{4\pi}\int d\vec r
d\vec r'dQ dQ'  n(\vec r,\vec Q)\frac{\vec Q\cdot \vec{Q}'}{|\vec
r-\vec r'|}n(\vec r',\vec Q')\Big)} \label{CANO}
\end{eqnarray}
The color charges are treated classically and we refer
to~\cite{johnson, litim&manuel, cho&zahed} for further details
regarding the nature of the measure. Here, we have defined

\begin{equation}
n(\vec r,\vec Q)=\sum_{i}^{N}\delta(\vec r-\vec r_i)\delta(\vec
Q-\vec Q_i) \label{eq002f}
\end{equation}
The generalization to many species is straightforward. The
associated Coulomb parameter is

\begin{equation}
\Gamma=\frac{g^2}{4\pi}\frac{\beta\,C_2}{a_{WS}} \label{GAMMA}
\end{equation}
where $C_2$ is the quadratic Casimir ($=\sum_i Q_i^2/(N_c^2-1)$)
and $a_{WS}$ is the Wigner-Seitz radius
${4\pi}a_{WS}^3/3=1/n$. For small $\Gamma$, Eq. (\ref{CANO})
behaves as a screened but weakly coupled gas, while for
intermediate values of $\Gamma$, Eq. (\ref{CANO}) describes a
liquid \cite{gelmanetal,gelmanetal2}. At large values of $\Gamma$
Eq. (\ref{CANO}) yields a solid as a ground state. From now on,
the canonical charge of $g^2/4\pi$ will be set to 1 for
simplicity, and will be restored in the final parameters by
inspection.

The static correlations both in space and in phase space
associated with Eq. (\ref{CANO}) are involved and will be the
subject of most of this paper. For that, we note that Eq.
(\ref{CANO}) yields the free energy generating functional

\begin{eqnarray}
\mathcal{F}_N[\psi]& &= \frac{1}{\beta}\int d\vec r dQ
n^{(1)}(\vec r,\vec Q)\Big(\ln{(\lambda^{3}n^{(1)}
(\vec r,\vec Q))}-1\Big) \nonumber \\
& &- \int d\vec r \psi(\vec r,\vec Q)n^{(1)}(\vec r,\vec
Q)+\mathcal{F}_{c}(n^{(1)}(\vec r,\vec Q))
\nonumber \\
& &+\frac{1}{2}\int d\vec r\vec r'dQdQ' n^{(1)}(\vec r,\vec
Q)\frac{\vec Q\cdot\vec{Q}'}{|\vec r-\vec r'|}n^{(1)}(\vec r',\vec
Q') \label{FREEDENSITY}
\end{eqnarray}
Here we have set

\begin{equation}
n^{(1)}(\vec r,\vec Q)=\langle n(\vec r,\vec
Q)\rangle=\langle\sum_{i}^{N}\delta(\vec r-\vec r_i)\delta(\vec
Q-\vec Q_i)\rangle \label{DENSITY}
\end{equation}
as the expectation value with the averaging carried using Eq.
(\ref{CANO}). The second contribution in Eq. (\ref{FREEDENSITY})
is the ideal classical contribution following from the measure in
Eq. (\ref{CANO}) using the asymptotic Stirling formulae. The third
contribution is the excess free energy functional. $\mathcal{F}_c$
is the connected free energy that sums up the second and higher
cumulants of $n(\vec r, \vec Q)$ from Eq. (\ref{CANO}). We note
that for zero scalar source $\psi=0$,

\begin{equation}
\mathcal{F}_N[0]=\mathcal{F}_{id}+\mathcal{F}_{ex} \label{EXCESS}
\end{equation}
where the first contribution is the classical ideal part and the
second contribution the excess part.

\section{Ornstein-Zernicke Equations}

\renewcommand{\theequation}{III.\arabic{equation}}
\setcounter{equation}{0}

To quantify the static interactions between pairs of particles in
(\ref{CANO}) we define

\begin{eqnarray}
-\frac{1}{\beta}\frac{\delta^2 \mathcal{F}_N}{\delta \psi\delta
\psi} & & = \Big<\sum_{i,j}\delta(\vec r-\vec r_i)\delta(\vec
r'-\vec r_j)\delta(\vec Q-\vec Q_i)\delta(\vec Q'-\vec Q_j)\Big>
\nonumber \\
& & =\Big(n^{(1)}(\vec r,\vec Q)n^{(1)}(\vec r',\vec Q')
{\bf{h}}(\vec r-\vec r',\vec Q\cdot\vec Q')+n^{(1)}(\vec r, \vec Q)
\delta(\vec r-\vec r')\delta(\vec Q-\vec Q')\Big) \nonumber \\
\label{HCORRELATOR}
\end{eqnarray}
with ${\bf h}$ the pair correlation function for $\psi=0$. The
pair correlation function is invariant under space translation and
color rotation. Generically

\begin{equation}
{\bf{h}}(\vec r-\vec r',\vec Q\cdot\vec Q')=\frac 1{n^2}
\Big<\sum_{i\neq j}\delta(\vec r-\vec r_i)\delta(\vec r'-\vec
r_j)\delta(\vec Q-\vec Q_i)\delta(\vec Q'-\vec Q_j)\Big>
\label{HCORRELATOR1}
\end{equation}
The direct correlation function ${\bf c}_D$ follows from the
excess free energy (\ref{EXCESS}) through

\begin{equation}
-\frac{1}{\beta}{\bf c}_D (\vec r-\vec r',\vec Q\cdot\vec
Q')=\frac{\delta^2\mathcal{F}^{ex}}{\delta n^{(1)}\delta
n^{(1)}}=\bigg(\frac{\vec Q\cdot\vec Q'}{|\vec r-\vec
r'|}+\frac{\delta^2 \mathcal{F}_{c}}{\delta
n^{(1)}\delta{n^{(1)}}}\bigg) \label{CD}
\end{equation}
which plays the role of a correlated potential. ${\bf c}_D$ will
be used below as a renormalized Coulomb potential in the liquid
phase. It also obeys the identity

\begin{equation}
\frac{\delta^2\mathcal{F}_N}{\delta n^{(1)}\delta n^{(1)}}=-\frac{
\delta\psi}{\delta
n^{(1)}}=\frac{1}{\beta}\bigg(\frac{1}{n}\delta(\vec r-\vec
r')\delta(\vec Q-\vec Q')-{\bf c}_D(\vec r-\vec r',\vec Q\cdot\vec
Q')\bigg) \label{CD1}
\end{equation}
Using the chain rule,

\begin{equation}
\int d\vec r''dQ''\frac{\delta\psi(\vec r,\vec Q)}{\delta
n^{(1)}(\vec r'',\vec Q'')}\frac{\delta n^{(1)}(\vec r'',\vec
Q'')}{\delta \psi(\vec r',\vec Q')} =\delta(\vec r-\vec
r')\delta(\vec Q-\vec Q') \label{eq010r}
\end{equation}
we obtain

\begin{equation}
{\bf h}(\vec r-\vec r',\vec Q\cdot\vec Q')={\bf c}_D (\vec r-\vec
r',\vec Q\cdot\vec Q') +n\int d\vec r''dQ''{\bf h}(\vec r-\vec
r'',\vec Q\cdot\vec Q''){\bf c}_D(\vec r''-\vec r',\vec
Q'\cdot\vec Q'') \label{OZ}
\end{equation}
which is the Orstein-Zernicke equation that ties the pair
correlation ${\bf h}$ to the direct correlation or the pair
potential ${\bf c}_D$. For a uniform plasma (\ref{OZ}) unfolds
agebraically in momentum and color space using

\begin{eqnarray}
& &{\bf h}(\vec r-\vec r',\vec Q\cdot\vec Q')=\int d\vec k
e^{i\vec k\cdot(\vec r-\vec r')}\sum_l \frac{2l+1}{4\pi}{\bf
h}_l(\vec k)P_l(\vec Q\cdot\vec Q') \nonumber \\
& &{\bf c}_D(\vec r-\vec r',\vec Q\cdot\vec Q')=\int d\vec k
e^{i\vec k\cdot(\vec r-\vec r')}\sum_l \frac{2l+1}{4\pi} {\bf
c}_{Dl}(\vec k)P_l(\vec Q\cdot\vec Q') \label{OZ1}
\end{eqnarray}
Thus

\begin{equation}
{\bf h}_l(\vec k)={\bf c}_{Dl}(\vec k)+n{\bf h}_{l}(\vec k)\,{\bf
c}_{Dl}(\vec k) \label{OZ2}
\end{equation}
which holds for each partial waves. (\ref{OZ2}) are the
generalized Orstein-Zernicke equations for each color partial wave
of the SU(2) colored Coulomb plasma.

\section{Static Structure Factors}

\renewcommand{\theequation}{IV.\arabic{equation}}
\setcounter{equation}{0}

The statistical aspects of the colored charged particles are best
captured by correlations in the phase space distributions. The
static structure factor is defined as

\begin{equation}
{\bf S}_0(\vec r-\vec r',\vec p \vec p',\vec Q\cdot \vec
Q')=\langle \delta f(\vec r\vec p \vec Q)\delta f(\vec r'\vec
p'\vec Q')\rangle \label{S1}
\end{equation}
with formally

\begin{equation}
f(\vec r\vec p\vec Q)=\sum_i\delta(\vec r-\vec x_i)\delta(\vec
p-\vec p_i)\delta(\vec Q-\vec Q_i) \label{PHASE}
\end{equation}
and $\delta f=f-\langle f\rangle$. The averaging in (\ref{S1}) is
carried using the canonical partition function (\ref{CANO}). Color
and translational invariance imply

\begin{equation}
\langle f(\vec r\vec p\vec Q) \rangle= nf_0(\vec
p)=n\Big(\frac{\beta}{2\pi m}\Big)^{3/2}e^{-\beta\vec p^2/2m}
\label{MEAN}
\end{equation}
which is the Maxwellian distribution for massive constituent
quarks or gluons. It is readily shown that

\begin{equation}
{\bf S}_0(\vec r-\vec r',\vec p\vec p',\vec Q\cdot\vec Q
')=nf_0(\vec p)\delta(\vec r-\vec r')\delta(\vec p-\vec p')\delta(
\vec Q-\vec Q')+n^2f_0(\vec p)f_0(\vec p'){\bf h}(\vec r-\vec
r',\vec Q\cdot\vec Q') \label{S2}
\end{equation}
The reduced static structure factor

\begin{equation}
{\bf S}_0(\vec k,\vec Q\cdot\vec Q')=\frac{1}{n}\int d\vec p d\vec
p'\int d\vec k e^{i\vec k\cdot(\vec r-\vec r')} {\bf S}_0(\vec
r-\vec r',\vec p\vec p',\vec Q\cdot\vec Q')
\end{equation}
ties with the pair correlation function (\ref{HCORRELATOR1})
through

\begin{equation}
{\bf S}_0(\vec k,\vec Q\cdot \vec Q')=\delta (\vec Q-\vec
Q')+n{\bf h}(\vec k, \vec Q\cdot \vec Q')
\end{equation}
Its Legendre transform in the color charge reads

\begin{equation}
{\bf S}_{0l}(\vec k)=1+n{\bf h}_l(\vec k) \label{S3}
\end{equation}
So the knowledge of the partial-wave structure factor $S_l(k)$
yields the pair correlation ${\bf c}_{Dl}$ through (\ref{OZ1}) and
(\ref{S3})

\begin{equation}
1={\bf S}_{0l}(\vec k)^{-1}+n {\bf c}_{Dl}(\vec k) \label{S4}
\end{equation}

We note that in configuration space the lth partial wave of the
static structure factor is

\begin{equation}
{\bf S}_{0l}(\vec r-\vec{r}')= \frac  1n \int
d\vec{p}\,d\vec{p}'\, \int dQ\,P_l(\vec Q\cdot \vec Q')\, {\bf
S}_0(\vec r-\vec{r}', \vec{p}\vec{p}',\vec Q\cdot\vec Q')
\label{XD1}
\end{equation}
Using (\ref{HCORRELATOR1})  and (\ref{S2}) and enforcing space
translational and color rotational invariance in the averaging
process yields

\begin{eqnarray}
{\bf S}_{0l}(\vec r)=\delta (\vec r) +\frac 1N \Big<\sum_{i\neq j}
\delta(\vec{r}-\vec{r}_{ij})\,P_l(\vec Q_i\cdot\vec Q_j)\Big>
\label{XD2}
\end{eqnarray}
In particular, the two lowest static structure factors are the
density structure factor

\begin{eqnarray}
{\bf S}_{00}(\vec r)=\delta(\vec r)+\frac 1N \Big<\sum_{i\neq j}
\delta(\vec{r}-\vec{r}_{ij})\Big>=\delta(\vec r )+n{\bf h}_0(\vec
r) \label{XD3}
\end{eqnarray}
and the charge structure factor

\begin{eqnarray}
{\bf S}_{01}(\vec r)=\delta(\vec r)+\frac 1N \Big<\sum_{i\neq j}
\delta(\vec{r}-\vec{r}_{ij})\,\vec Q_i\cdot\vec Q_j\Big>
\label{XD4}
\end{eqnarray}
Higher structure factors are given by (\ref{XD2}) as they measure
the various color correlation content of the SU(2) strongly
coupled QGP. Below, we propose both an analytical and numerical
derivation of the two lowest structure factors (\ref{XD3}) and
(\ref{XD4}).

\section{Debye-Huckel-Hole Potential}

\renewcommand{\theequation}{V.\arabic{equation}}
\setcounter{equation}{0}

To derive the static structure factors we will use the Debye
charging procedure for a fixed color charge. For that, we need the
Poisson-Boltzman equation for the 1-species SU(2) colored plasma
in the presence of a colored test charge $\vec
q$~\cite{gelmanetal}

\begin{equation}
\nabla^2\vec \phi(\vec r,\vec r',\vec q)=-4\pi\Bigg(\vec
q\delta(\vec r-\vec r')+ \int dQ'\vec Q' n(\vec r,\vec q)
e^{-\beta\vec Q'\cdot(\vec \phi(\vec r,\vec r',\vec q)-\vec
\Phi(\vec r,\vec q))} \Bigg) \label{PB1}
\end{equation}
with the {\it fixed} external density profile

\begin{equation}
n(\vec r, \vec q)=n+n(\triangle_0+\vec \Delta_1\cdot\vec q)
\cos(\vec k\cdot\vec r) \label{PROFILE}
\end{equation}
We note that (\ref{PROFILE}) is a scalar under rigid and
orthogonal color rotations ${\bf R}Q$ if the external parameters
$\Delta_l$ transform unitarily as $D({\bf R})\Delta$ with $D({\bf
R})$ the Wigner rotation in the adjoint representation. This fixed
density causes an imposed potential

\begin{equation}
\nabla^2\vec\Phi(\vec r,\vec q)=-4\pi\,\vec q\,(n(\vec r,\vec
q)-n)
\end{equation}
which is used to normalize the Poisson-Boltzman equation in
(\ref{PB1}). We solve (\ref{PB1}) in the linear approximation. For
that we define the shifted potential $\delta\vec \phi=\vec
\phi-\vec \Phi$,

\begin{equation}
\Big(\nabla^2-\kappa_D^2 n(\vec r,\vec q)/n\Big)\delta \vec
\phi(\vec r,\vec r',\vec q)\approx-4\pi\vec q\Big(\delta(\vec
r-\vec r')-(n(\vec r,\vec q)-n)\Big) \label{PB2}
\end{equation}
with $\kappa_D^2\equiv 4\pi\beta nC_2$ the squared Debye constant.
(\ref{PB2}) is the linearized Poisson-Boltzman or Debye-Huckel
equation for the classical colored plasma. At short separations
(\ref{PB2}) is known to misrepresent the hole caused by the strong
Coulomb correlations. To fix that we use the Debye-Huckel plus
hole approximation\cite{nordholm}

\begin{equation}
\Big(\nabla^2-\kappa_D^2 n(\vec r,\vec q)/n\Theta\Big)\delta \vec
\phi(\vec r,\vec r',\vec q)\approx-4\pi\vec q\Big(\delta(\vec
r-\vec r')-(n(\vec r,\vec q)-n)\Big)(1-\Theta) \label{PB3}
\end{equation}
with $\Theta=\theta(|\vec r-\vec r'|-\sigma)$ the spherical hole
insertion of radius $\sigma$. The mean-induced potential is

\begin{equation}
\vec\Psi(\vec r,\vec q)=\lim_{\vec r\rightarrow \vec
r'}\Big(\delta\vec \phi(\vec r,\vec r',\vec q)-\frac{\vec q}{|\vec
r-\vec r'|}\Big) \label{SCREENINGPOTENTIAL}
\end{equation}

\section{Debye Charging Process}

\renewcommand{\theequation}{VI.\arabic{equation}}
\setcounter{equation}{0}

To assess the static structure factors for the classical and
strongly coupled colored plasma, we note that the excess free
energy (\ref{EXCESS}) can be readily rewritten in terms of the
pair correlations

\begin{equation}
-\beta\mathcal{F}^{ex}=\frac{1}{2}\int d\vec r d\vec r'dQdQ'
n^{(1)}(\vec r,\vec Q)c_D(\vec r-\vec r',\vec Q \cdot\vec Q')
n^{(1)}(\vec r',\vec Q') \label{DR1}
\end{equation}
In Fourier (space)  and Legendre (color) space (\ref{DR1}) reads

\begin{equation}
-\beta\mathcal{F}^{ex}=\frac{1}{2}\int d\vec{k}\, dQ\,dQ'
\sum_{l,m}n^{(1)}_{lm}(\vec k,\vec Q)\,{\bf c}_{Dl}(\vec k )\,
n^{(1)}_{lm}(-\vec k ,\vec Q') \label{DR2}
\end{equation}
with

\begin{equation}
n^{(1)}_{lm}(\vec k,\vec Q)=\int d\vec r e^{-i\vec k\cdot\vec
r}Y_l^m(\vec Q)n^{(1)}(\vec r,\vec Q) \label{eq019}
\end{equation}
and $Y_l^m$ a spherical harmonic for an SU(2) colored plasma.
Using the partial wave form of the Orstein-Zernicke equations
(\ref{S4}), (\ref{DR2}) becomes

\begin{equation}
-\beta\mathcal{F}^{ex}=\frac{1}{2n}\int d\vec k\, dQ\, dQ'
\sum_{l,m} n^{(1)}_{lm}(\vec k,\vec Q)\Big(1-{\bf
S}_{0l}^{-1}(\vec k)\Big) n^{(1)}_{lm}(-\vec k,\vec Q')
\label{DR3}
\end{equation}
This shows that the quadratic change in the excess free energy
caused by an external density profile $n_{lm}(k)$ is directly
proportional to the $lth$ partial wave of the inverse of the
static structure factor.

The external density profile (\ref{PROFILE}) changes the color
Coulomb potential locally through (\ref{PB2}), thereby affecting
the free energy. To assess the change in the latter we use the
Debye charging procedure~\cite{mcquarrie}. For that, we note that
by dialing (\ref{PROFILE}) the free energy shifts. The shift can
be decomposed into three parts,

\begin{equation}
\mathcal{F}=\mathcal{F}_{\rm ideal}+\mathcal{F}_{\rm
imposed}+\mathcal{F}_{\rm induced} \label{SPLIT}
\end{equation}
The shift in the ideal part is set by the first term in
(\ref{FREEDENSITY}) after inserting (\ref{PROFILE}). The imposed
contribution is

\begin{equation}
\mathcal{F}_{\rm imposed}=\int d\vec r dQ (n(\vec r,\vec q)-n)\vec
q\cdot \int_{0}^{1}d\lambda\vec \Phi(\vec r,\lambda\vec q)
\label{IMPOSED}
\end{equation}
and follows from the imposed charge. Specifically,

\begin{equation}
\mathcal{F}_{\rm imposed}=\frac{n^2}{2}q^2(3\Delta_0^2+|\vec
\Delta_1|^2) \int d\vec r d\vec r'\frac{\cos(\vec k\cdot\vec
r)\cos(\vec k\cdot\vec r')}{|\vec r-\vec r'|} \label{IMPOSED1}
\end{equation}

The induced free energy is

\begin{equation}
\mathcal{F}_{\rm induced}=\int d\vec r dQ  n(\vec r,\vec q)\vec
q\cdot \int_{0}^{1} d\lambda\vec \Psi(\vec r,\lambda\vec q)
\label{INDUCED}
\end{equation}
and follows from the induced but shifted screening potential
(\ref{SCREENINGPOTENTIAL}).

\begin{figure}[!h]
\begin{center}
\includegraphics[width=0.495\textwidth]{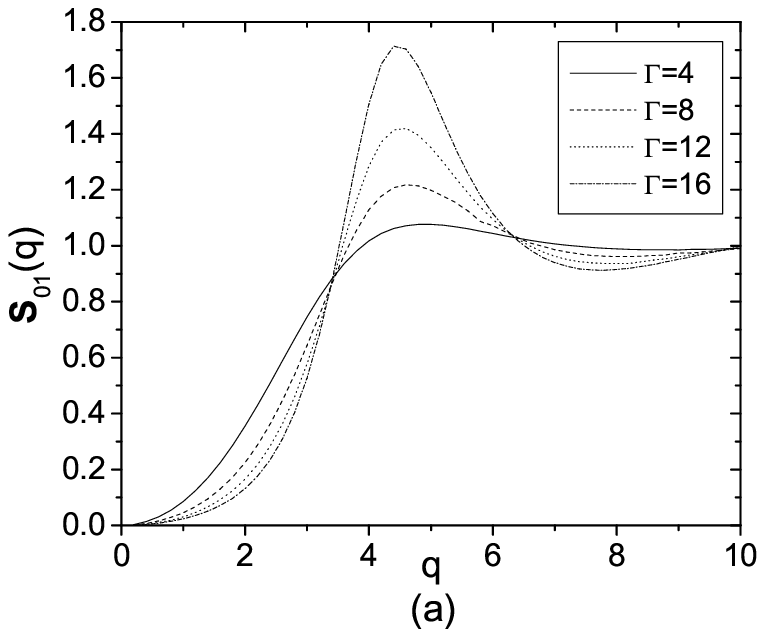}
\includegraphics[width=0.495\textwidth]{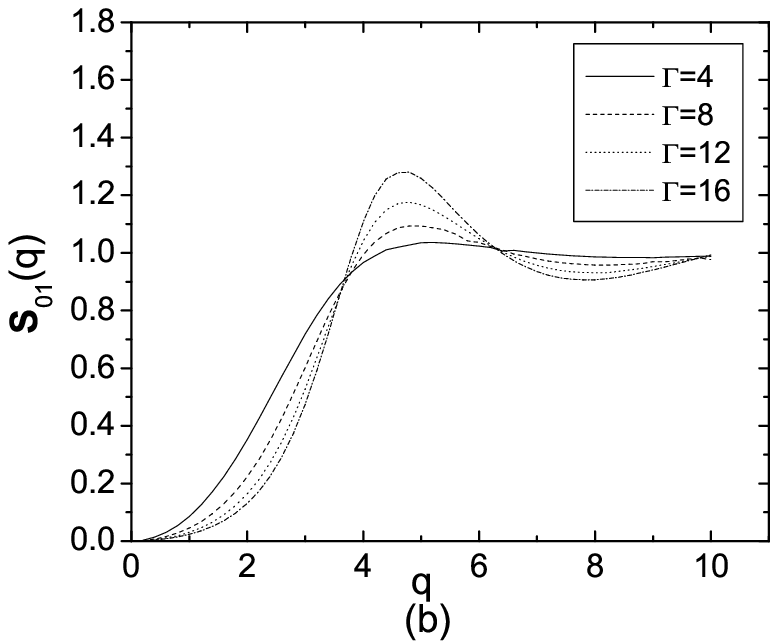}
\end{center}
\caption{${\bf S}_{01} (q)$ for $\Gamma=4,8,12,16$ summed up to
$l=1$ (a) and $l=2$ (b). See text.}
\end{figure}
The integrand can be obtained by solving (\ref{PB3}) for
$\delta\vec \phi(\vec r,\vec r',\vec q)$ with the help of the
Green function,

\begin{equation}
\bigg[\nabla^2-\kappa_D^2\Theta(|\vec r-\vec
r'|-\sigma)\bigg]G(\vec r'', \vec r-\vec r')=-4\pi\delta^3(\vec
r''-(\vec r-\vec r')) \label{eq020d}
\end{equation}
The method has been developed
in~\cite{lee&fisher1,lee&fisher2,tamashiroetal} for the
one-component plasma and readily extends to our colored plasma.
For that we evaluate the reduced free energy $f=\beta {\cal F}/V$
to quadratic order in $\Delta^2$. By comparing the terms with
(\ref{DR3}), we can extract ${\bf S}_{00}^{-1}$ as the coefficient
of $\Delta^2_0$ and ${\bf S}_{01}^{-1}$ as the coefficient of
${\vec\Delta}^2_1$.   We find that both static structure factors
are finite and identical in this approximation,

\begin{eqnarray}
& &  {\bf S}_{00}^{-1}(\vec k)={\bf S}_{01}^{-1}(\vec k)=
1+2(\frac{\kappa_D}{k})^2 \int_{0}^{1}d\lambda
\frac{\lambda}{w_{\lambda}}\cos{(\frac{1}{\lambda}\frac{k}{\kappa_D}
(w_{\lambda}-1))} \nonumber \\
& & \quad \qquad +2 (\frac{\kappa_D}{k})^3\int_{0}^{1}d\lambda
\frac{\lambda^2}{w_{\lambda}}\sin{(\frac{1}{\lambda}\frac{k}{\kappa_D}
(w_{\lambda}-1))}-2\int_{0}^{1}\frac{d\lambda} {\lambda}
\frac{w_{\lambda}-1}{w_{\lambda}}\mathcal{I}_{0}^{+}
(w_{\lambda}-1,\frac{1}{\lambda}\frac{k}{\kappa_D})
\nonumber \\
& & \quad \qquad
+\sum_{l=0}^{\infty}(2l+1)\bigg(2\frac{\kappa_D}{k} \int_{0}^{1}
d\lambda \frac{(w_{\lambda}-1)^{l+2}}
{w_{\lambda}g_{l+1}(w_{\lambda}-1)} j_{l+1}(\frac{1}{\lambda}
\frac{\kappa_D}{k} (w_{\lambda}-1))
\mathcal{I}_{l}^{+}(w_{\lambda}-1,\frac{1}{\lambda}\frac{k}{\kappa_D})
\nonumber \\
& & \qquad \quad +(-1)^{l+1}\int_{0}^{1} \frac{d\lambda}
{\lambda}\frac{w_{\lambda}-1}{w_{\lambda}}
\Big(\frac{g_{l+1}(-w_{\lambda}+1)}{g_{l+1}(w_{\lambda}-1)}
(\mathcal{I}_{l}^{+}(w_{\lambda}-1,\frac{1}
{\lambda}\frac{k}{\kappa_D}))^2 \nonumber \\
& & \qquad \quad +2\mathcal{I}_{l}^{+} (w_{\lambda}-1,\frac{1}
{\lambda}\frac{k}{\kappa_D}) \mathcal{I}_{l}^{-}
(w_{\lambda}-1,\frac{1} {\lambda}\frac{k}{\kappa_D})
-2\mathcal{I}_{l}^{0} (w_{\lambda}-1,\frac{1}
{\lambda}\frac{k}{\kappa_D})\Big) \bigg) \label{eq021d}
\end{eqnarray}
The three integral contributions are

\begin{eqnarray}
& & \mathcal{I}_{l}^{-}(z, y)=\int_{0}^{z}dw
w^{-l}g_l(-w)j_l(y w) \nonumber \\
& & \mathcal{I}_{l}^{0}(x, y)=\int_{x}^{\infty}dz
z^{-l}g_l(z)j_l(y z)\mathcal{I}_{l}^{-}(z, y)e^{(2(x-z))} \nonumber \\
& & \mathcal{I}_{l}^{+}(x, y)=\int_{x}^{\infty}dz
z^{-l}g_l(z)j_l(y z)e^{(2(x-z))} \label{eq022d}
\end{eqnarray}
with $g_{l}(z) =e^{z}z^{l+1}k_{l}(z)$. Here $j_{l}(yz)$ is a
spherical Bessel function, and $k_{l}(z)$ a modified spherical
Bessel function.  The parameter $w_{\lambda}$ is defined as

\begin{equation}
w_{\lambda}=\Big(1+\lambda^3(3\Gamma)^{\frac{3}{2}}\Big)^{\frac{1}{3}}
\label{eq023d}
\end{equation}
The static structure factors in (\ref{eq021d}) involve summations
over multiple partial waves. The sums are rapidly converging as we
show in  Fig.~1 with $l=1$ retained (a) and $l=2$ retained (b).
Here $q=ka_{WS}$ is a dimensionless wave-vector. To assess the
accuracy of the analytical method developed above for the static
correlation functions in the colored Coulomb plasma, we now carry
numerical simulations for the same structure factors using
molecular dynamics simulations.

\section{Structure Factors from Molecular Dynamics}

\renewcommand{\theequation}{VII. \arabic{equation}}
\setcounter{equation}{0}

For an SU(2) plasma, the details of the molecular dynamics
simulations can be found in~\cite{gelmanetal}. Color motion is
treated as a point coordinate on a 3-sphere with a fixed radius
that is equal to the quadratic Casimir for SU(2). Classical
stability of the colored Coulomb gas at short distances is
achieved by using a scalar core potential of the type

\begin{equation}
V_{core}=\frac{1}{n}\frac{1}{|\vec r_i -\vec r_j|^n}
\label{eq001l}
\end{equation}
with $n=9$.  The two-body interparticle colored potential
is~\cite{gelmanetal,hansen&mcdonald4},

\begin{equation}
V(r,
\vec{Q}\cdot\vec{Q}')=\frac{g^2}{\lambda}\Big[\frac{1}{9}\Big(\frac{\lambda}{r}\Big)^9
+\vec Q\cdot\vec Q'\Big(\frac{\lambda}{r}\Big)\Big] \label{eq002l}
\end{equation}
with $\lambda$ setting the unit of length scale.  At close packing
the density is $n_{\rm cp}=1/\lambda^3$. We choose the unit of
length $\lambda$ so that $n_{\rm cp}=1$. The unit of time is set
by the inverse plasma frequency $\tau=\omega_p^{-1}$. In these
units, the strength of the colored Coulomb potential is
$\frac{1}{4\pi}\frac{1}{n\lambda^3}$ \cite{gelmanetal}.

We have adopted the Verlet algorithm in~\cite{hansen&mcdonald} to
integrate the equations of motion for a system composed of 108
particles. The particles are confined in a box and surrounded by
images via periodic boundary conditions. The simulations are
carried in a fixed volume $1/(n\lambda^3)=V/(N\lambda^3)=2.72$ as
in~\cite{hansen&mcdonald4}. The Wigner-Seitz radius $a_{WS}=(4\pi
n/3)^{-1/3}$ is $0.866\lambda$. With these parameters, the
interparticle interaction strength is set by the Coulomb constant
$\Gamma$.

\begin{figure}[!h]
\begin{center}
\subfigure{\label{radial:all}\includegraphics[width=0.50\textwidth]
{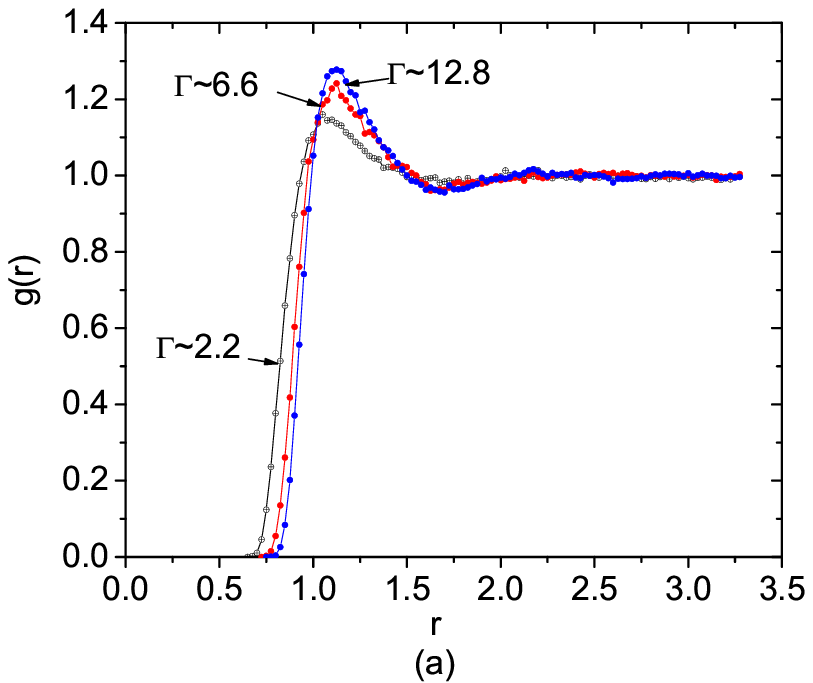}}
\subfigure{\label{radial:G021}\includegraphics[width=0.49\textwidth]
{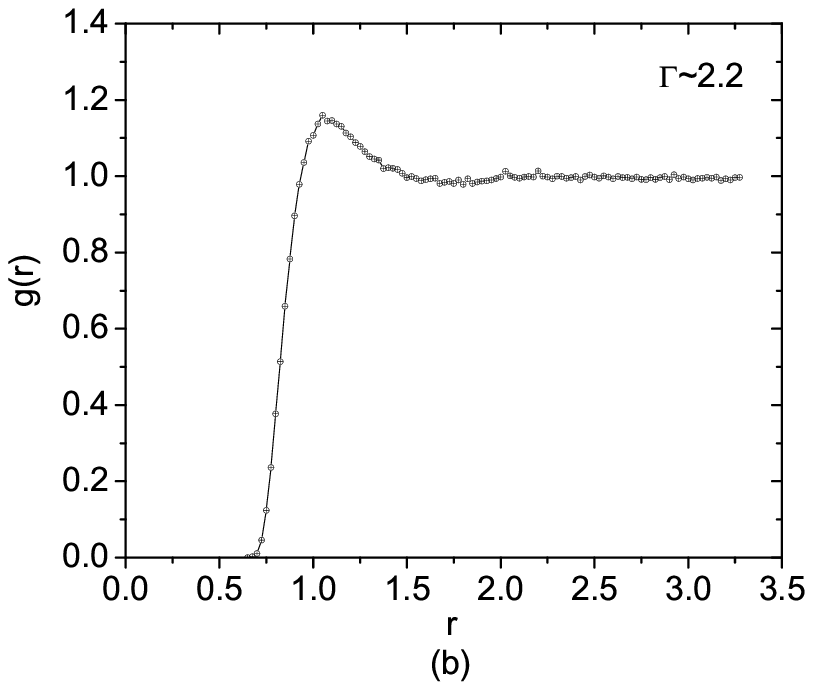}}
\end{center}
\begin{center}
\subfigure{\label{radial:G066}\includegraphics[width=0.50\textwidth]
{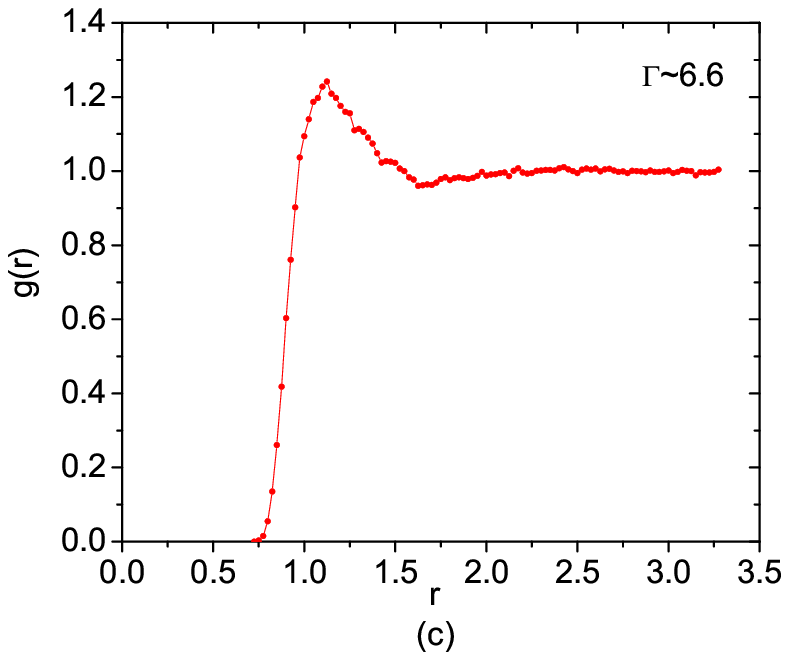}}
\subfigure{\label{radial:G128}\includegraphics[width=0.49\textwidth]
{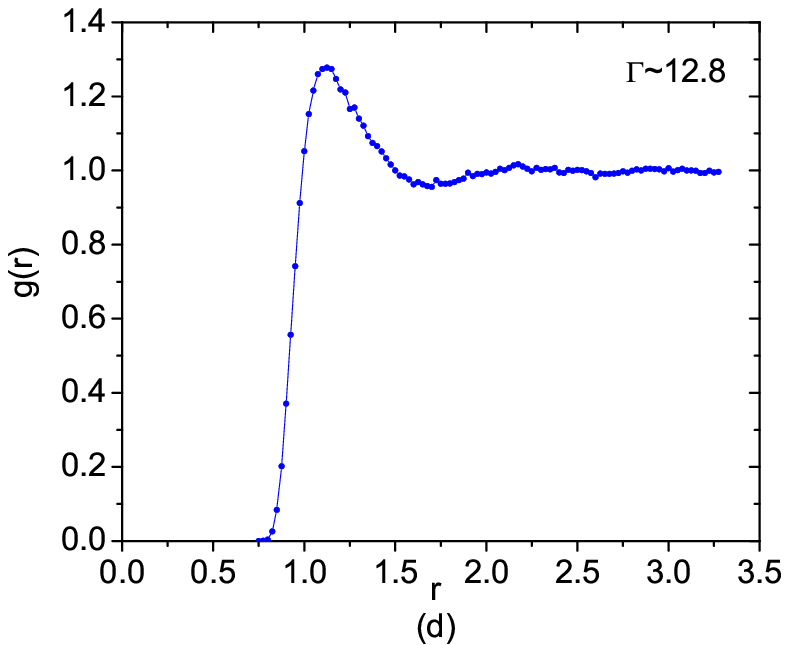}}
\end{center}
\caption{Radial distribution function for $\Gamma =$2.2(b) ,
6.6(c) and 12.8(d). See text.} \label{radial}
\end{figure}

We first measure the particle radial distribution function ${\bf
g}(r)={\bf h}_0(r)$ as a function of $\Gamma$.  ${\bf g}(r)$
measures the probability of finding two particles between $r$ and
$r+\Delta{r}$,

\begin{equation}
{\bf g}(r)\equiv {\bf h}_0(r)=\frac{1}{nN}\Big< \sum_{i\neq j}^{N}
\delta(\vec r-\vec r_{ij})\Big> \label{eq3I}
\end{equation}
In Fig.~\ref{radial} we show ${\bf g}(r)$ versus $r$ for different
Coulomb couplings $\Gamma=2.2, 6.6, 12.8$. The larger $\Gamma$ the
larger the size of the Coulomb hole surrounding each colored
Coulomb particle. Also, the larger $\Gamma$, the higher the peak,
the tighter the Coulomb packing.

\begin{figure}[!h]
\begin{center}
\includegraphics[width=0.49\textwidth]{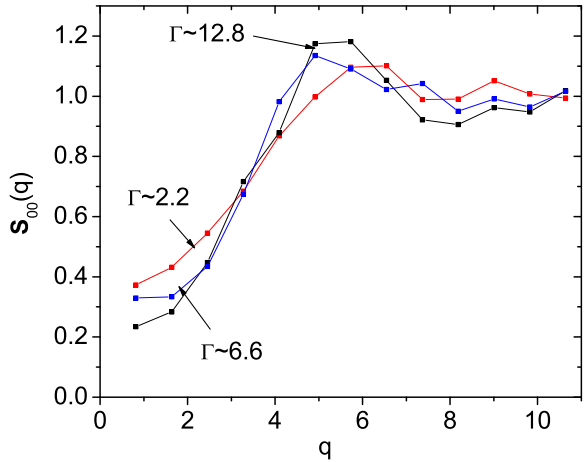}
\end{center}
\caption{Static Structure factor ${\bf S}_{00} (q)$  versus $q$
for $\Gamma=2.2, 6.6, 12.8$. See text.} \label{structure0}
\end{figure}

The radial distribution function in Fig.~\ref{radial} for the
SU(2) colored Coulomb plasma appears overall similar to the one
observed for the one component plasma (OCP).  Although our colored
particles attract for color antiparallell charges, they overall
statistically repel due to the larger color repulsive
orientations. The difference with the OCP is best seen by taking
the Fourier transform of (\ref{eq3I}) which is the $l=0$ density
structure factor ${\bf S}_{00}(q)$

\begin{equation}
{\bf S}_{00}(\vec k)=\frac{1}{N}\langle |{\bf n}_{\vec
k}|^2\rangle \label{eq004l}
\end{equation}
with

\begin{equation}
{\bf n}_{\vec k}=\sum_{i=1}^Ne^{i\vec{k}\cdot\vec{r}_i}
\end{equation}
Fig.~\ref{structure0} shows the behavior of ${\bf S}_{00}(q)$
versus the dimensionless wave-vector $q=a_{WS}k$.  The
nonvanishing  of ${\bf S}_{00}$ at the origin reflects on the
coupling to the sound mode. In the static and long-wavelength
approximation it is just

\begin{equation}
{\bf S}_{00}(\vec k)\approx \frac{\vec{k}^2}{c_S^2\vec
k^2}=\frac 1{c_S^2}
\end{equation}
with $c_S^2=(\partial P/\partial \rho)_T$ the isothermal squared
speed of sound, with $P$ the pressure and $\rho$ the mass density.
Since $k$ is a multiple of $2\pi /L$ because of the finite cubic
box $L\times L\times L$, only about a dozen points were accessible
numerically. Since $L\approx N^{1/3}$, we need to increase the
number of particles in the box to smoothen out the structure
factor in momentum space.

\begin{figure}[!h]
\begin{center}
\subfigure{\label{structure:one}\includegraphics[width=0.495\textwidth]
{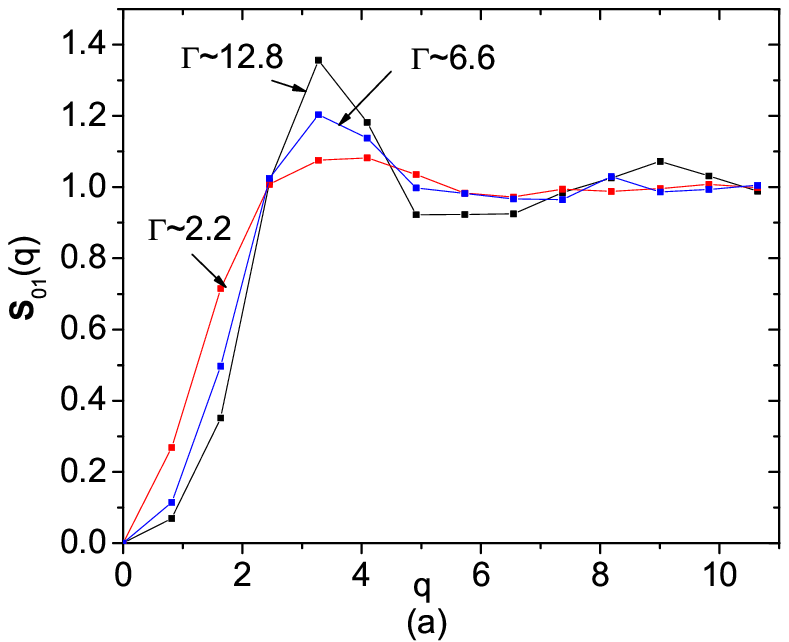}}
\subfigure{\label{structure:G021}\includegraphics[width=0.495\textwidth]
{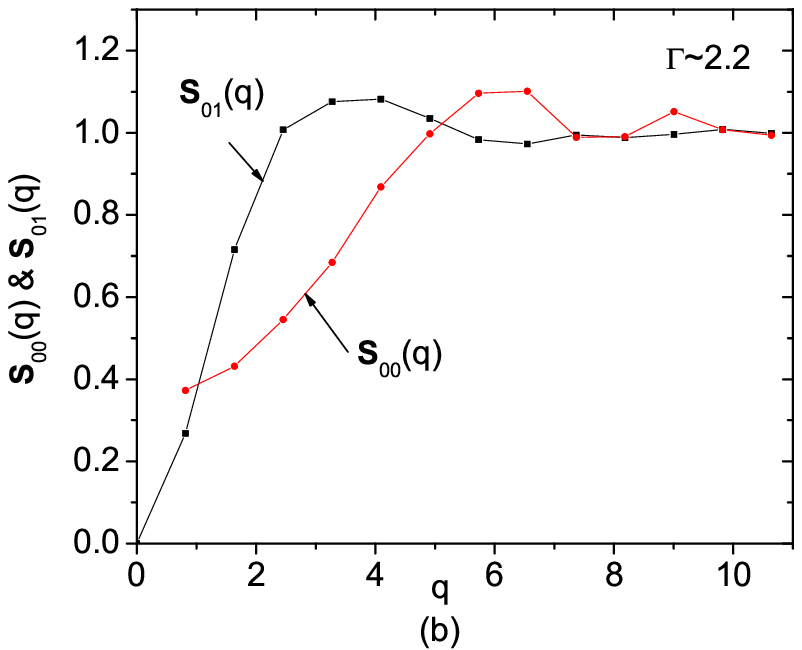}}
\end{center}
\begin{center}
\subfigure{\label{structure:G066}\includegraphics[width=0.495\textwidth]
{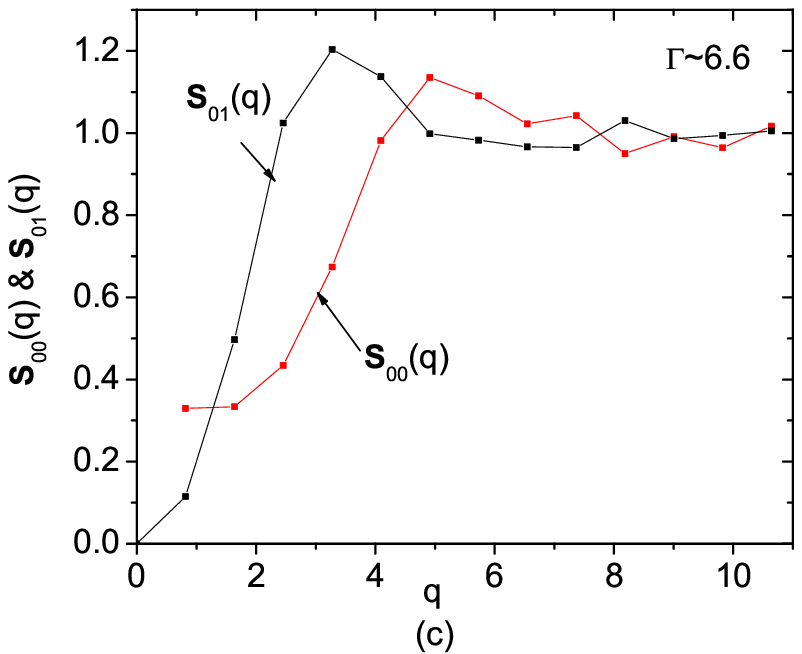}}
\subfigure{\label{structure:G128}\includegraphics[width=0.495\textwidth]
{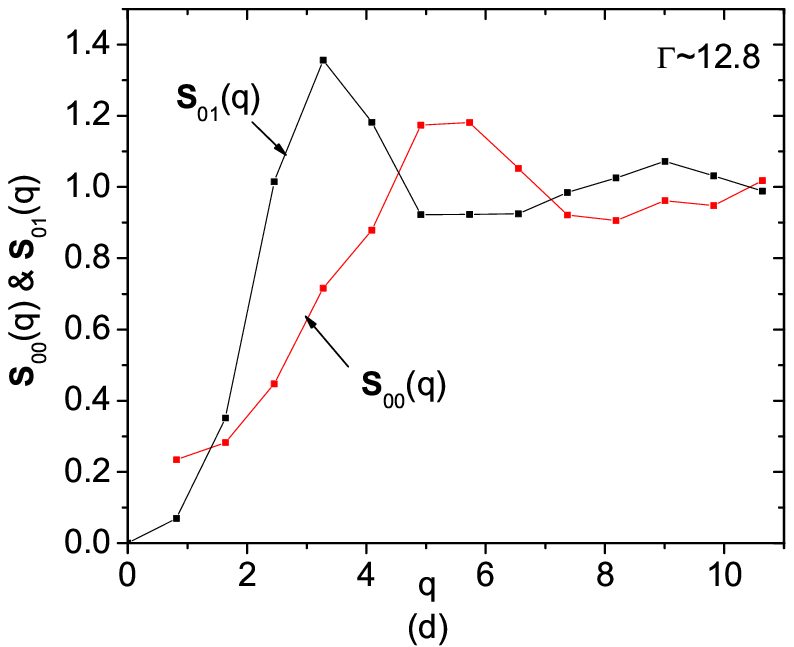}}
\end{center}
\caption{The static structure factor ${\bf S}_{01}(q)$ for
$\Gamma=2.2, 6.6, 12.8$. See text.}\label{structure}
\end{figure}

In Fig.~\ref{structure}, we show the $l=1$ or charge  structure
factor

\begin{equation}
{\bf S}_{01}(k)=\frac{1}{N}\langle|\vec\rho_{\vec k}|^2\rangle
\label{eq5I}
\end{equation}
with

\begin{eqnarray}
{\vec \rho}_{\vec k}=\sum_{i=1}^N\,\vec
Q_i\,e^{i\vec{k}\cdot\vec{r}_i}
\end{eqnarray}
Unlike ${\bf S}_{00}$ which correlates a pair of scalar densities,
${\bf S}_{01}$ correlates a pair of charge densities. In the OCP
plasma both correlators are identical. They are not in the SU(2)
colored Coulomb plasma. In the long wavelength approximation, the
static density structure factor is saturated by the plasmon mode

\begin{equation}
{\bf S}_{01}(\vec k)\approx \frac{\vec{k}^2}{k_D^2}
\end{equation} which is seen to vanish at zero momentum.

Our analytical result for ${\bf S}_{01}(k)$ in (\ref{eq021d})  is
in agreement with the molecular dynamics simulations for the
charged correlator  (\ref{eq5I}). Our analytical result for ${\bf
S}_{00}(k)$ is identical with ${\bf S}_{01}(k)$. It differs from
the molecular dynamics simulation results for small momenta since
the sound mode drops out of the Debye-Huckel colored potential on
which our charging process was based. The contribution of the
sound mode is additive at small momenta, and drops out at large
momentum due to  damping through the shear viscosity.

\section{Conclusions}

\renewcommand{\theequation}{IX.\arabic{equation}}
\setcounter{equation}{0}

The strongly coupled SU(2) QGP is characterized by a number of
static correlators in phase space with color treated as a
classical 3-vector on $S^3$ with a radius fixed by the second
Casimir. Space translational invariance and color rotational
invariance yields  multiple structure factors characterizing
color correlations with color charges sourced by Legendre
polynomials. Each structure factor obeys a generalized
Ornstein-Zernicke equation.

To evaluate analytically these multiple structure factors, we have
made use of the Debye charging process and the linearized
Poisson-Boltzman equation in line with linear response theory. We
have derived explicit relations for the two lowest structure
factors, ie $l=0,1$ which corresponds to the density and charge
structure factors.

To check the validity of the linear response analysis, we have
numerically extracted the density and charge static structure
factors using SU(2) molecular dynamics simulations. Modulo the
sound mode, both analytical structure factors compare favorably
with the numerical results. The current analysis extends to higher
multipoles, ie $l=2,3,...$ and generalizes to higher color
SU(${\rm N}>2$) groups.

The static structure factors play an important role in
characterizing the correlations in the colored SU(2) QGP at
intermediate and large values of the coupling coupling $\Gamma$.
They also enter in the assessment of transport parameters at
strong coupling. The results will be presented elsewhere.

The current classical and strongly coupled SU(2) colored Coulomb
plasma can be extended to several species to account for gluons,
quarks and antiquarks~\cite{gelmanetal}.  The effects of quantum
mechanics being a renormalization of the constituent
parameters such as the mass and charge. It will be interesting to
see whether a quantum phase space formulation of QCD is achievable
through the background field formulation in a way that allows for
the introduction of colored static structure factors.

\section{Acknowledgments}

We thank Kevin Dusling for discussions. This work was supported in
part by US-DOE grants DE-FG02-88ER40388 and DE-FG03-97ER4014.

\appendix

\section{SU(2) Color charges}

\renewcommand{\theequation}{A.\arabic{equation}}
\setcounter{equation}{0}

The explicit representation of the classical color charges is~\cite{johnson,litim&manuel}

\begin{equation}
Q^1=\cos\phi_1 \sqrt{J^2-\pi_1^2}, \quad  Q^2=\sin\phi_1
\sqrt{J^2-\pi_1^2}, \quad Q^3=\pi_1 \label{eq001aa}
\end{equation}
with $J^2$ the quadratic Casimir $q_2=
\sum_{\alpha}^{N_c^2-1}{Q^{\alpha}Q^{\alpha}}$. The measure in the SU(2) phase space can be set to

\begin{equation}
dQ=c_R d\pi_1 d\phi_1 J dJ \delta(J^2-q_2) \label{eq002aa}
\end{equation}
where $c_R$ is a representation dependent constant. These SU(2)
color charges satisfy

\begin{eqnarray}
& & \int dQ Q^{\alpha}=0 \nonumber \\
& & \int dQ Q^{\alpha}Q^{\beta}=C_2\delta^{\alpha\beta}
\label{eq003aa}
\end{eqnarray}
For fixed Casimir $\sum_{\alpha}Q^{\alpha}Q^{\alpha}
=(N_c^2-1)C_2$,  we can chose the spherical representation
for (\ref{eq001aa})

\begin{equation}
Q^1=\sin{\theta}\cos{\phi}, \quad Q^2=\sin{\theta}\sin{\phi},
\quad Q^3=\cos{\theta} \label{eq004aa}
\end{equation}
for which the measure (\ref{eq002aa}) reads

\begin{equation}
dQ=\sin{\theta}d\theta d\phi \label{eq005aa}
\end{equation}
Equivalently,

\begin{eqnarray}
& & Q^1=-\sqrt{\frac{2\pi}{3}}\Big(Y_1^{-1}(\theta,\phi)-Y_1^1(
\theta,\phi)\Big) \nonumber \\
& & Q^2=i\sqrt{\frac{2\pi}{3}}\Big(Y_1^{-1}(\theta,\phi)+Y_1^1(
\theta,\phi)\Big) \nonumber \\
& & Q^3=\sqrt{\frac{4\pi}{3}}Y_1^0(\theta,\phi) \label{eq006aa}
\end{eqnarray}
in terms of spherical harmonics. In the spherical representation,
we have $\int dQ=4\pi$,
$\sum_{\alpha}Q^{\alpha}Q^{\alpha} =1$ and $\int dQ \vec
Q\cdot\vec Q=4\pi$.


\end{document}